\documentclass[preprint]{aastex}

\usepackage{graphicx}

\begin{document}

\title{Simulating the Smallest Ring World of Chariklo}
\author{Shugo Michikoshi, and Eiichiro Kokubo}
\altaffiltext{1}{ Center for Computational Sciences, University of Tsukuba, Tsukuba, Ibaraki 305-8577, Japan }
\altaffiltext{2}{
  Division of Theoretical Astronomy, National Astronomical Observatory of
  Japan, Osawa, Mitaka, Tokyo 181-8588, Japan
}

\email{ michikos@ccs.tsukuba.ac.jp, and kokubo@th.nao.ac.jp }

\begin{abstract}
  
A ring system consisting of two dense narrow rings has been discovered
 around Centaur Chariklo. 
The existence of these rings around a small object poses various questions,
 such as their origin, stability, and lifetime. 
In order to understand the nature of Chariklo's rings, we perform
  global $N$-body simulations of the self-gravitating collisional particle rings for the first time. 
We find that Chariklo should be denser than the ring material to avoid
 the rapid diffusion of the rings. 
If Chariklo is denser than the ring material, fine spiral structures
 called self-gravity wakes occur in the inner ring. 
These wakes accelerate the viscous spreading of the ring significantly and they typically occur
 on timescales of about $100\,\mathrm{years}$ for m-sized ring
 particles, which is considerably shorter than the timescales suggested in previous
 studies.   
The existence of these narrow rings implies smaller ring particles or
the existence of shepherding satellites.  

\end{abstract}

\section{Introduction}

Recently, two dense narrow rings around Centaur Chariklo were discovered
 by occultation observation \citep{Braga-Ribas2014}.
 A ring around Centaur Chiron was also proposed \citep{Ruprecht2015, Ortiz2015}. 
These observations suggest that rings around large Centaurs may not be as rare as previously thought. 

Several mechanisms for the formation of Chariklo's rings have been proposed:
 collisional ejection from the parent body, satellite disruption, and out
 gassing \citep{Pan2016}.
The tidal disruption of a differentiated object during a close encounter
 with a giant planet can also result in the formation of a ring \citep{Hyodo2016}. 
From the numerical integration of the Chariklo orbit it was suggested that the possibility of a close encounter with a giant planet that is able to disrupt Chariklo's rings is low \citep{Araujo2016}. 
The formation mechanism of the rings is still not well understood. 
In order to understand the origin of the ring system, we need to investigate the ring structure in detail. 

Observations have found a difference in the inner ring widths between
 ingress and egress \citep{Braga-Ribas2014}.
The width variation indicates that the inner ring has a finite eccentricity. 
On the other hand, Chariklo has large oblateness, which should lead to 
the rapid differential precession of the ring, resulting in a circular ring.
Thus, there must be some mechanism keeping the ring eccentricity. 
To explain the eccentric ring around Uranus, apse alignment due
 to ring self-gravity has been proposed \citep{Goldreich1979b, Goldreich1979a}. 
By applying the same model to Chariklo's rings, the ring mass and
 particle size have been estimated \citep{Pan2016}.
From these estimates, they found that the inner ring is slightly
 gravitationally unstable in the context of Toomre stability \citep{Toomre1964}. 

Local $N$-body simulations of Saturn's rings revealed that
 self-gravity wakes exist in Saturn's A and B rings \citep[e.g.,][]{Salo1992a, Salo1995, Michikoshi2011, Michikoshi2015}.
The self-gravity wakes are non-axisymmetric small spiral structures
 formed by gravitational instability, which enable efficient angular
 momentum transfer.
Thus, it is important to elucidate whether self-gravity wakes exist in Chariklo's rings.
The optical depth in Chariklo's inner ring is about $0.38$, which is
 marginal for the formation of self-gravity wakes \citep{Salo1995}.
The critical optical depth for the formation of self-gravity wakes depends on the
 distance from the central body and the properties of the ring particles. 
In order to investigate Chariklo's ring system, we perform
 global $N$-body simulations of the two narrow rings with full
 self-gravity for the first time. 
 Section \ref{sec:sim} describes the model and the simulation method,
 Section \ref{sec:res} presents the results of the numerical simulations, and
 Section \ref{sec:summary} contains a summary and a discussion.

\section{Simulation Method \label{sec:sim}}

Chariklo has a large oblateness of about $0.2$ \citep{Braga-Ribas2014}, 
resulting in a large gravitational moment $J_2$, which
 causes the differential precession of the rings.
We here focus on the formation of structures over the dynamical timescale.
Thus, we neglect the oblateness. 
The latest observations and analysis show that Chariklo has a shape with volume equivalent to a sphere of radius 
$125\,\mathrm{km}$ and its density is $0.8$ or $1.2\, \mathrm{g}\,\mathrm{cm}^{-3}$ \citep{Leiva2016}.
We adopt a spherical Chariklo with radius $125\,\mathrm{km}$ and density $\rho_\mathrm{C} = 1.0\, \mathrm{g}\,\mathrm{cm}^{-3}$.  
Chariklo has the inner dense ring,
with different optical depths and radial widths for the ingress ($W=6.16\, \mathrm{km}$, $\tau=0.449$) and egress
 ($W=7.17\, \mathrm{km}$, $\tau=0.317$). 
This indicates that the inner ring may have a finite eccentricity.
However, in this paper, we assume that the inner ring is circular and
 uniform with $W=6.7\, \mathrm{km}$ and $\tau=0.38$ as a first step. 
Similarly, we assume that the outer ring is circular and uniform
 with $W=3.5\, \mathrm{km}$ and $\tau=0.06$.
The inner and outer rings are located at distances of $a = 390.6\,\mathrm{km}$ and
 $404.8\,\mathrm{km}$ from the center of Chariklo, respectively. 

The properties of ring particles are not well constrained.
For the sake of simplicity, we assume that all particles have the same mass $m_\mathrm{p}$ and radius $r_\mathrm{p}$. 
In Saturn's rings, the typical particle size varies between a centimeter and a few meters \citep{French2000}.  
Furthermore, considering the apse alignment of the ring by its self-gravity,
 \cite{Pan2016} estimated the ring mass, which corresponds to a few
 meter-sized particles.   
  We assume $r_\mathrm{p} = 2.5 \mbox{--} 10\, \mathrm{m}$, and density ratios $\rho_\mathrm{p}/\rho_\mathrm{C} = 0.05, 0.10, 0.25, 0.50, 0.75,$ and $1.00$. 
The number of ring particles is $21$ to $345$ million.
The basic ring dynamics is controlled by the two non-dimensional ring
 parameters $\tau$ and $\tilde r_\mathrm{H}$, where 
 $\tilde r_\mathrm{H}$ is the scaled Hill radius of ring particles given
 by $\tilde r_\mathrm{H} = r_\mathrm{H}/2 r_\mathrm{p}$ 
 with $r_\mathrm{H} = (2m_\mathrm{p}/3M_\mathrm{C})^{1/3} a$ \citep{Salo1995}, where $M_\mathrm{C}$ is Chariklo's mass.
Note that $\tilde r_\mathrm{H}$ is independent of the particle size. 

The equation of motion of particle $i$ is
\begin{equation}
 \frac{\mathrm{d}^2 \mathbf{r}_i}{\mathrm{d}t^2} =
 - G M_\mathrm{C} \frac{\mathbf r_i}{|\mathbf r_i|^3}
 - \sum_{j\ne i} G m_\mathrm{p}
  \frac{\mathbf{r}_{ij}}{(r_{ij}^2 + \epsilon_\mathrm{g}^2)^{3/2}}
 + \mathbf{f}_\mathrm{col},
\end{equation}
 where $\mathbf{r}_i$ is the position vector of particle $i$, 
 $\mathbf{r}_{ij} = \mathbf{r}_{i} - \mathbf{r}_{j}$,
 $r_{ij}= |\mathbf{r}_{ij} |$, $\epsilon_\mathrm{g}$ is the softening length, and $\mathbf{f}_\mathrm{col}$ is the collisional force.
We adopt the soft sphere model as the collision model \citep{Salo1995}.
In this model, a collision is described as a damped oscillation. 
This model has a free parameter, the collision duration
 $t_\mathrm{col}$, which should be sufficiently shorter than the dynamical timescale.
The collision duration is $t_\mathrm{col} = 0.0025 t_\mathrm{K}$, where $t_\mathrm{K}$ is the orbital period. 
The restitution coefficient is $\epsilon = 0.1$. 
We set $\epsilon_\mathrm{g} = 0.1 r_\mathrm{p}$.

We integrate the equation of motion with the leapfrog method.
To reduce the computational time, as in \cite{Salo1995}, we consider
 two different timesteps, $\Delta t_\mathrm{grav}$ and
 $\Delta t_\mathrm{col}$.
The timestep $\Delta t_\mathrm{col}$ is for collisions, 
and we integrate the equation of motion with $\Delta t_\mathrm{col}$. 
We update the gravitational acceleration with $\Delta t_\mathrm{grav}$,
 which is set to be longer than $\Delta t_\mathrm{col}$.
We adopt $\Delta t_\mathrm{grav} = 0.005 t_\mathrm{K}$, which is sufficiently
 small for resolving gravitational interactions among particles, and
 $\Delta t_\mathrm{col} = \Delta t_\mathrm{grav}/32 = t_\mathrm{col}/16$. 
 We use the tree algorithm for the gravitational force calculation \citep{Barnes1986}.  
We adopt the opening angle $\theta = 0.5$, and
employ the $N$-body simulation library, Framework for Developing Particle Simulator (FDPS)
\citep{Iwasawa2016} with the Phantom-GRAPE module \citep{Tanikawa2012, Tanikawa2013}.

\section{Results \label{sec:res}}

\subsection{Formation of Self-gravity Wakes}

\begin{figure}
\begin{center}
\includegraphics[width=\hsize]{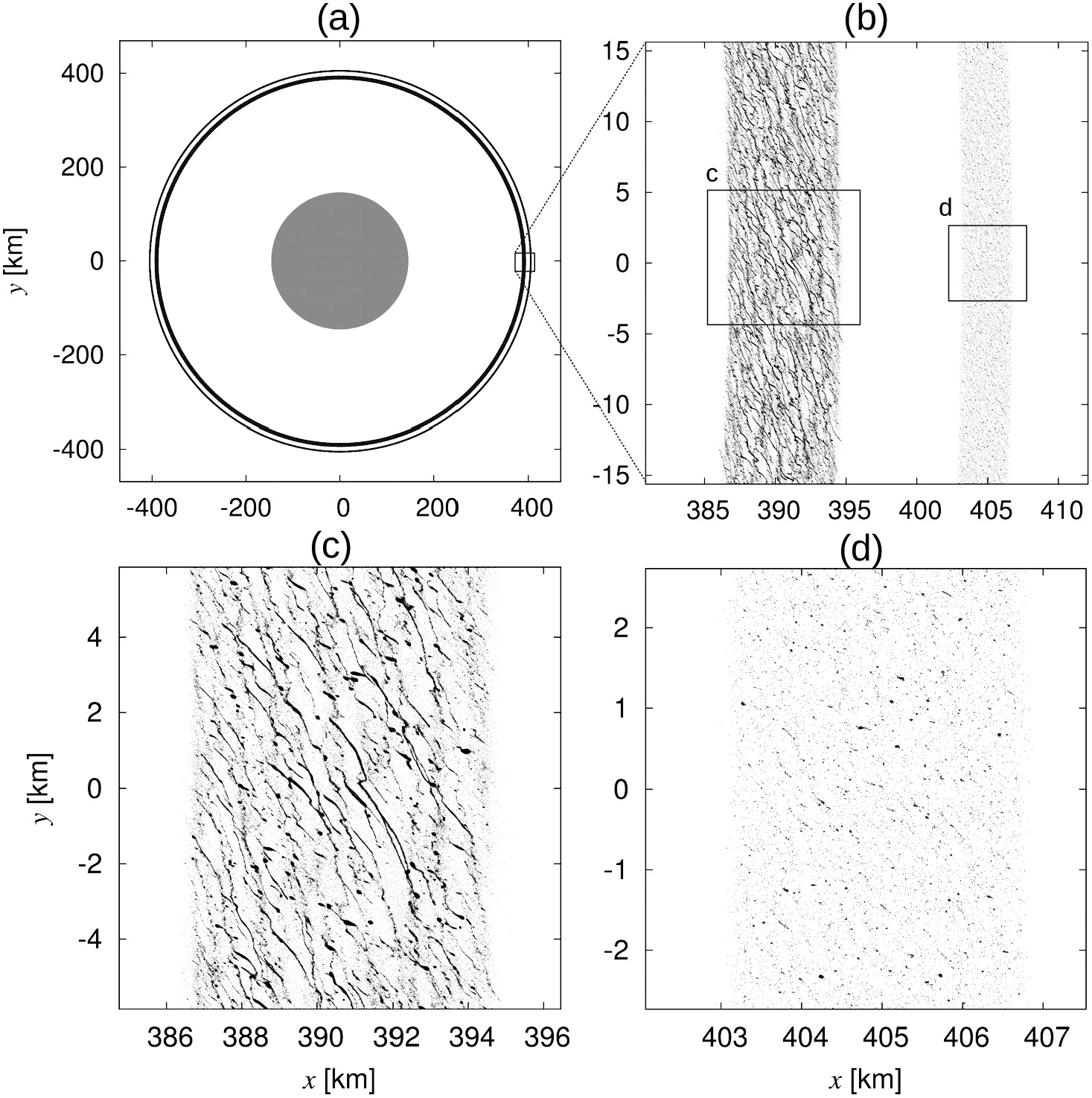}
\end{center}
\caption{
  Snapshots of the simulated ring in the $x$--$y$ plane at $t = 10  t_\mathrm{K}$, where
  the particle density and radius are $\rho_\mathrm{p}/\rho_\mathrm{C} = 0.5$ and $r_\mathrm{p} = 5\,\mathrm{m}$,
 respectively. 
The left panel (a) shows the overall structure of the ring, while the panels (b), (c), (d) show enlarged views of ring sections.
\label{eq:fid_model}
}
\end{figure}

\begin{figure}
\begin{minipage}{0.5\hsize}
\begin{center}
(a) $t = 0 t_\mathrm{K}$\\
\includegraphics[width=\hsize]{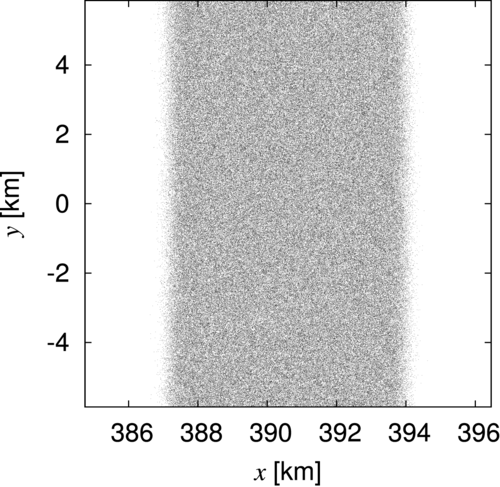}
\end{center}
\end{minipage}
\begin{minipage}{0.5\hsize}
\begin{center}
(b) $t = 1 t_\mathrm{K}$\\
\includegraphics[width=\hsize]{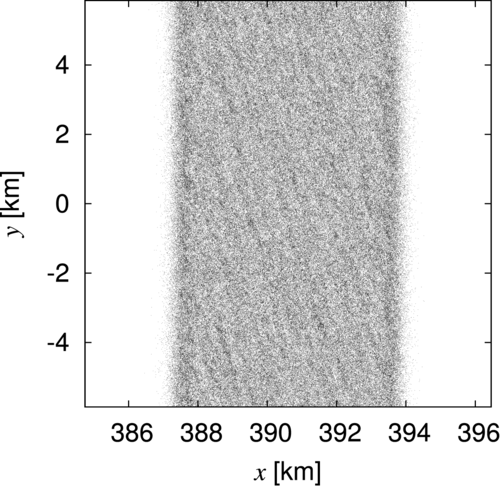}
\end{center}
\end{minipage}
\begin{minipage}{0.5\hsize}
\begin{center}
(c) $t = 2 t_\mathrm{K}$\\
\includegraphics[width=\hsize]{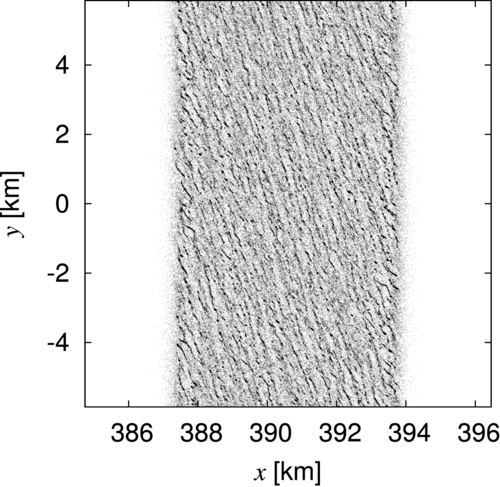}
\end{center}
\end{minipage}
\begin{minipage}{0.5\hsize}
\begin{center}
(d) $t = 3 t_\mathrm{K}$\\
\includegraphics[width=\hsize]{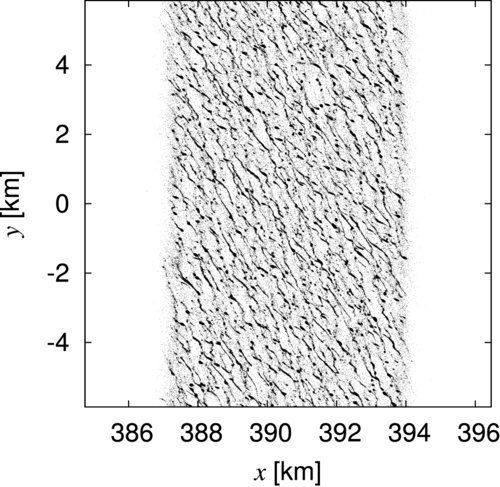}
\end{center}
\end{minipage}
\caption{
  Snapshots of the inner ring at (a) $t = 0$, (b) $1 t_\mathrm{K}$, (c) $2 t_\mathrm{K}$, and (d) $3 t_\mathrm{K}$ for
  the model with $\rho_\mathrm{p}/\rho_\mathrm{C} = 0.5$ and $r_\mathrm{p} = 5\,\mathrm{m}$.
\label{fig:evolution}
}
\end{figure}

Figure \ref{eq:fid_model} shows the ring at
 $t = 10 t_\mathrm{K}$, where $\rho_\mathrm{p}/\rho_\mathrm{C} = 0.50$ 
 and $r_\mathrm{p}=5\,\mathrm{m}$;
no large-scale structures are visible.
For dynamical timescales, the ring keeps its original global
 shape.
However, we find small-scale structures, namely, self-gravity wakes, in the inner ring
(Fig. \ref{fig:evolution}).
Initially, the ring particles are distributed uniformly.
At $t = 1 t_\mathrm{K}$, fluctuations in the ring surface density grow
 due to gravitational instability, and 
self-gravity wakes form at $t = 2 t_\mathrm{K}$.
After $t = 3 t_\mathrm{K}$, the self-gravity wakes are formed and
 destroyed continuously. 
The self-gravity and collision of particles form particle aggregates,
 while the differential rotation tears them apart.
These competing processes are the cause of self-gravity wakes.

The spatial scale of self-gravity wakes is characterized by the critical
 wavelength of the gravitational instability 
 \citep{Toomre1964, Salo1995, Daisaka1999}
\begin{equation}
\lambda_\mathrm{cr} =
\frac{4 \pi^2 G\Sigma}{\Omega^2} =
0.36 \left(\frac{a_i}{390.6 \, \mathrm{km}} \right)^3
 \left(\frac{r_\mathrm{C}}{125 \, \mathrm{km}} \right)^{-3}
  \left(\frac{\rho_\mathrm{p}/\rho_\mathrm{C}}{0.5}\right)
 \left(\frac{\tau}{0.38}\right)
 \left(\frac{r_\mathrm{p}}{5 \, \mathrm{m}}\right) \, \mathrm{km},  
\label{eq:lambda}
\end{equation}
 where $\Sigma$ is the ring surface density and
 $\Omega = 2 \pi /t_\mathrm{K}$ is the Kepler angular frequency.
 
On the other hand, in the outer ring, the self-gravity wakes do not develop.
Since the optical depth in the outer ring is small, the energy
 dissipation by collisions is insufficient.
Thus, the random velocity of ring particles is high and the ring is
 gravitationally stable.
Since the outer ring is near the Roche limit, small aggregates are visible. 
The tidal force is comparable with their self-gravity. Thus they may be temporal.

\subsection{Ring Particle Properties}

\begin{figure}
\begin{minipage}{0.5\hsize}
\begin{center}
(a) $\rho_\mathrm{p} /\rho_\mathrm{C} = 0.05$\\
\includegraphics[width=0.85\hsize]{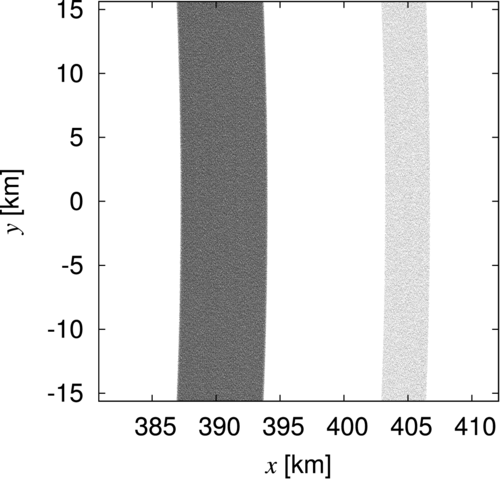}
\end{center}
\end{minipage}
\begin{minipage}{0.5\hsize}
\begin{center}
(b) $\rho_\mathrm{p} /\rho_\mathrm{C} = 0.10$\\
\includegraphics[width=0.85\hsize]{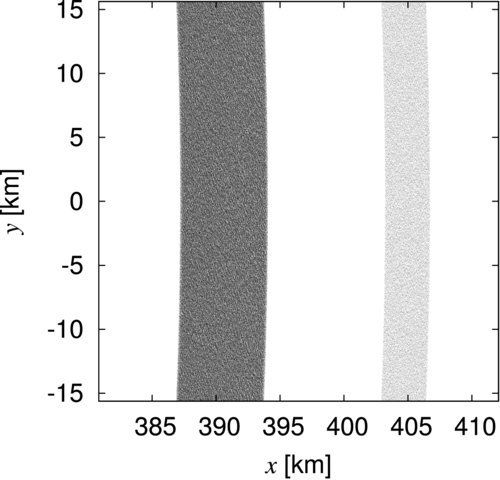}
\end{center}
\end{minipage}
\begin{minipage}{0.5\hsize}
\begin{center}
(c) $\rho_\mathrm{p} /\rho_\mathrm{C} = 0.25$\\
\includegraphics[width=0.85\hsize]{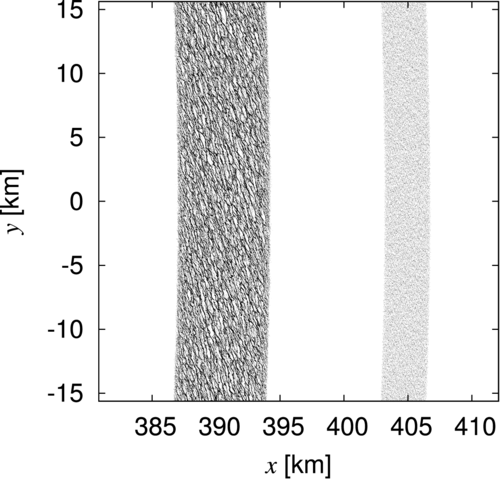}
\end{center}
\end{minipage}
\begin{minipage}{0.5\hsize}
\begin{center}
(d) $\rho_\mathrm{p} /\rho_\mathrm{C} = 0.50$\\
\includegraphics[width=0.85\hsize]{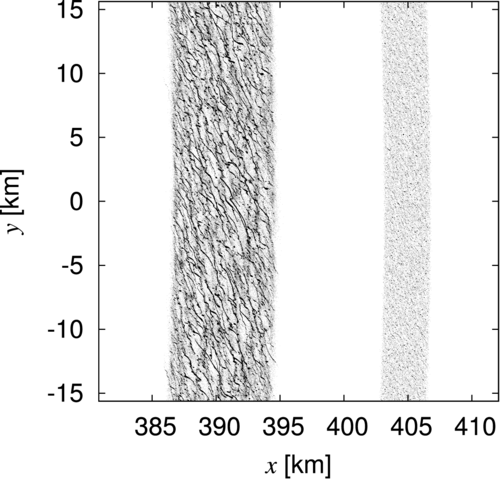}
\end{center}
\end{minipage}
\begin{minipage}{0.5\hsize}
\begin{center}
(e) $\rho_\mathrm{p} /\rho_\mathrm{C} = 0.75$\\
\includegraphics[width=0.85\hsize]{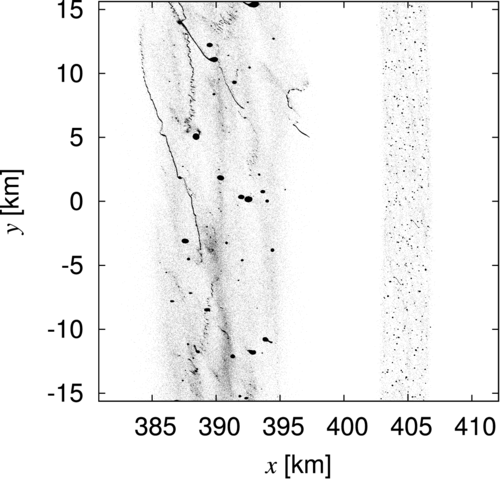}
\end{center}
\end{minipage}
\begin{minipage}{0.5\hsize}
\begin{center}
(f) $\rho_\mathrm{p} /\rho_\mathrm{C} = 1.00$\\
\includegraphics[width=0.85\hsize]{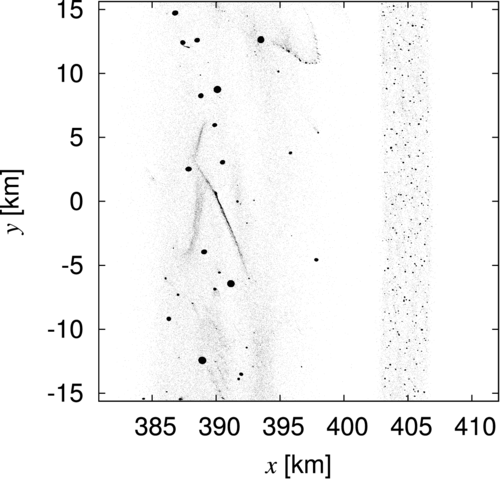}
\end{center}
\end{minipage}
\caption{
  Snapshots of the inner ring at $t=10 t_\mathrm{K}$ for the models
  with (a) $\rho_\mathrm{p} /\rho_\mathrm{C} = 0.05$, (b) $0.1$, (c) $0.25$, (d) $0.50$, (e) $0.75$, and (f) $1.0$ with $r_\mathrm{p} = 5 \, \mathrm{m}$.
}
\label{fig:rhopdep}
\end{figure}

We constrain the properties of ring particles with those of the
 self-gravity wakes.
Figure \ref{fig:rhopdep} shows the structure of the rings at
 $t = 10 \, t_\mathrm{K}$ for the models where
 $\rho_\mathrm{p}/\rho_\mathrm{C} =0.05, 0.10, 0.25, 0.50, 0.75$ and $1.00$ with $r_\mathrm{p} = 5\,\mathrm{m}$.
In the $\rho_\mathrm{p}/\rho_\mathrm{C} = 0.75,$ and $1.00$ models, large
 particle aggregates form and the ring spreads rapidly over the dynamical
 timescale.  
Thus, this model may not correspond to the real ring.
The scaled Hill radius determines whether a particle pair can
 gravitationally bind \citep{Ohtsuki1993a, Salo1995, Karjalainen2004}, which is given as
\begin{equation}
\tilde r_\mathrm{H} =
1.36 \left(\frac{a}{390.6\,\mathrm{km}}\right)
 \left(\frac{r_\mathrm{C}}{125\,\mathrm{km}}\right)^{-1}
 \left(\frac{\rho_\mathrm{p}}{\rho_\mathrm{C}}\right)^{1/3}.
\label{eq:rh}
\end{equation}
Studies of Saturn's rings showed that gravitationally bound aggregates
 form if $\tilde r_\mathrm{H} \gtrsim 1.1$ 
 \citep{Ohtsuki1993a, Salo1995, Karjalainen2004}. 
Therefore, the necessary condition for ring formation is
 $\rho_\mathrm{p}/\rho_\mathrm{C} < 0.52$, in other words, the particle density should be lower than half of that of Chariklo.

 In the inner ring, self-gravity wakes appear for $\rho_\mathrm{p}/\rho_\mathrm{C} = 0.25$ and $0.5$, while they do not appear for $\rho_\mathrm{p}/\rho_\mathrm{C} = 0.05$ and $0.1$.
As the particle density increases, the spatial scale of
 the self-gravity wakes increases, as shown in Equation (\ref{eq:lambda}).
For $\tau = 0.38$, the self-gravity wakes form under the condition
 $0.65 \lesssim \tilde r_\mathrm{H} \lesssim 1.1$
 \citep{Salo1995, Ohtsuki2000, Daisaka2001}, which is independent of the
 particle radius.  
If $\tilde r_\mathrm{H} \lesssim 0.65$, the self-gravity wake formation
 is suppressed, which corresponds to $\rho_\mathrm{p}/ \rho_\mathrm{C} < 0.11$.
If Chariklo has the typical Centaur density, $\rho_\mathrm{C} \simeq 1.0 \, \mathrm{g}\,\mathrm{cm}^{-3}$, the particle density needs to be less than $0.1 \, \mathrm{g}\,\mathrm{cm}^{-3}$ for the suppression of self-gravity wakes. 
 
\begin{figure}
\begin{minipage}{0.5\hsize}
\begin{center}
(a) $r_\mathrm{p} = 2.5\, \mathrm{m}$\\
\includegraphics[width=\hsize]{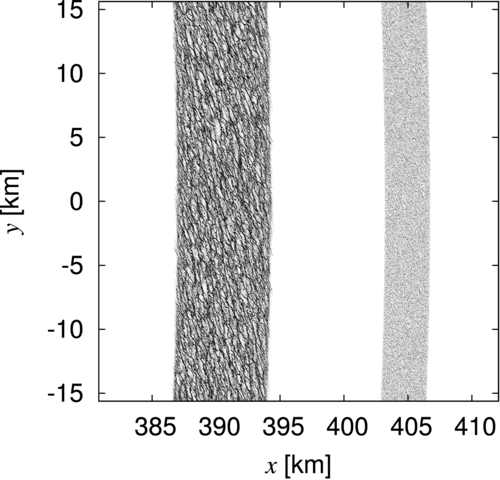}
\end{center}
\end{minipage}
\begin{minipage}{0.5\hsize}
\begin{center}
(b) $r_\mathrm{p} = 5\, \mathrm{m}$\\
\includegraphics[width=\hsize]{final3_rp0050_rho0050.png}
\end{center}
\end{minipage}
\begin{minipage}{0.5\hsize}
\begin{center}
(c) $r_\mathrm{p} = 7.5\, \mathrm{m}$\\
\includegraphics[width=\hsize]{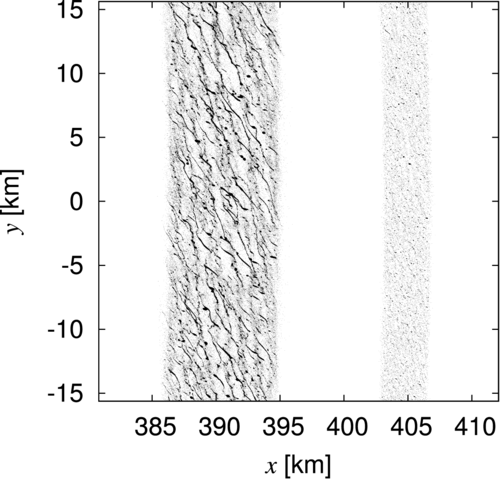}
\end{center}
\end{minipage}
\begin{minipage}{0.5\hsize}
\begin{center}
(d) $r_\mathrm{p} = 10\, \mathrm{m}$\\
\includegraphics[width=\hsize]{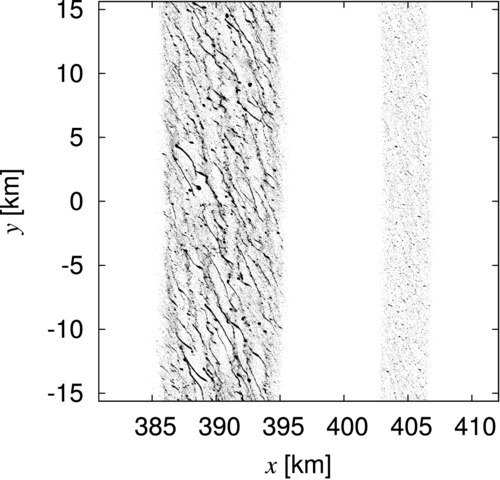}
\end{center}
\end{minipage}
\caption{
  Snapshots of the inner ring at $t=10 t_\mathrm{K}$ for the models
   with (a) $r_\mathrm{p} = 2.5 \, \mathrm{m}$, (b) $5 \, \mathrm{m}$, (c)
 $7.5 \, \mathrm{m}$, and (d) $10 \, \mathrm{m}$ with 
  $\rho_\mathrm{p} /\rho_\mathrm{C} = 0.50$.
}

\label{fig:rpdep}
\end{figure}

Figure \ref{fig:rpdep} shows the dependence of the spatial scale of 
self-gravity wakes on $r_\mathrm{p}$ with $\rho_\mathrm{p} / \rho_\mathrm{C} = 0.5$, 
$r_\mathrm{p} = 2.5, 5, 7.5,$ and $10 \, \mathrm{m}$.
In all the models, the self-gravity wakes develop, which is consistent
 with the condition for self-gravity wake formation discussed above.
As indicated by Equation (\ref{eq:lambda}), the spatial scale increases
 with $r_\mathrm{p}$.
The ring width for larger particles is wider than that with smaller particles.
This is because the viscosity due to self-gravity wakes increases with
 wake size. 

\begin{figure}
\begin{center}
\includegraphics[width=\hsize]{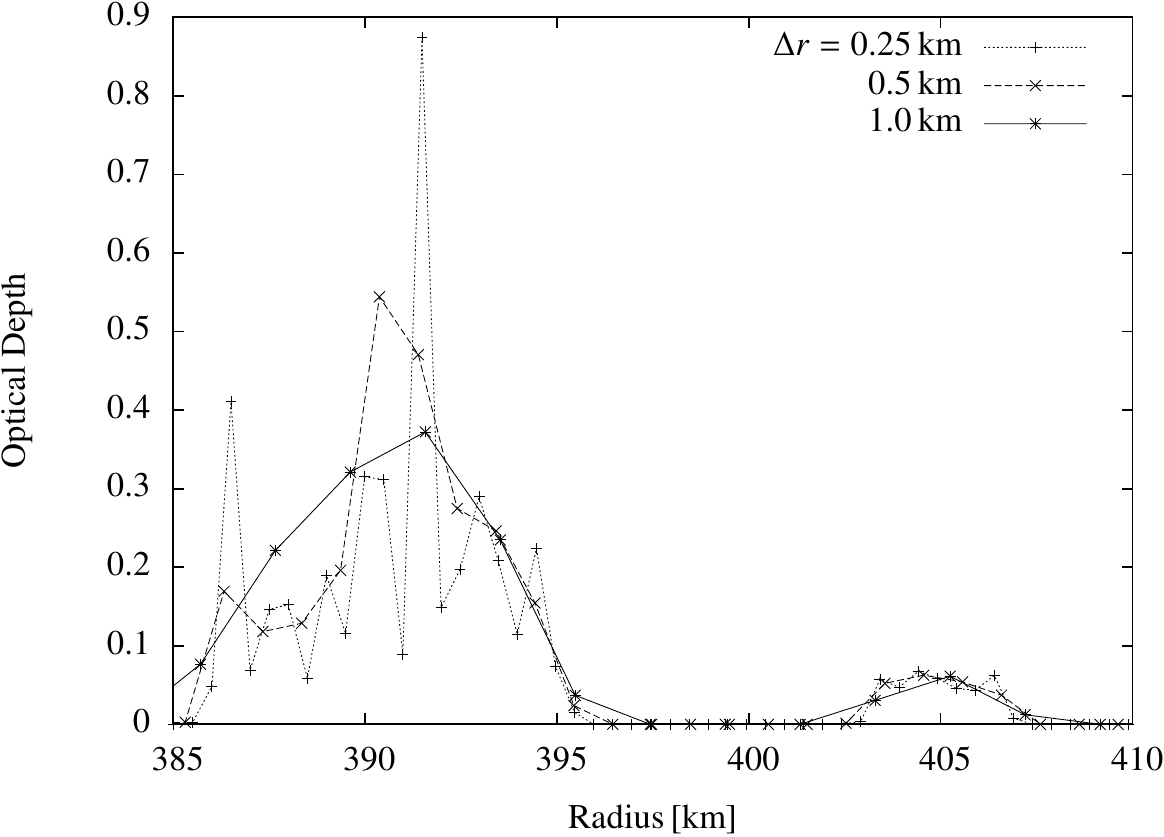}
\end{center}
\caption{
Simulation of the dynamical optical depth along the line
 $y = 0$ at $t = 10\,t_\mathrm{K}$ for the model where
 $\rho_\mathrm{p}/\rho_\mathrm{C} = 0.5$ and
 $r_\mathrm{p} = 10\,\mathrm{m}$.
  The spatial resolutions are $\Delta r=0.25\, \mathrm{km}$ (dotted), $\Delta r=0.5\, \mathrm{km}$ (dashed), and $1.0\, \mathrm{km}$ (solid).  \label{eq:optical_depth}
}  
\end{figure}

The Chariklo ring was discovered by stellar occultation
 \citep{Braga-Ribas2014}. 
The projected star radius for Chariklo was estimated at around $1\,\mathrm{km}$.
Figure \ref{eq:optical_depth} shows a simulation of the
 dynamical optical depth $\tau = 4 \Sigma / 3 r_\mathrm{p} \rho_\mathrm{p}$ with spatial resolutions with radii
 $\Delta r = 0.25$, $ 0.50$, and $1.0 \, \mathrm{km}$ for the model where
 $\rho_\mathrm{p} / \rho_\mathrm{C} = 0.5$ and
 $r_\mathrm{p} = 10 \, \mathrm{m}$.
 The critical wavelength is $\lambda_\mathrm{cr} = 0.72\, \mathrm{km}$.
We can observe fluctuations due to self-gravity wakes in the
 $\Delta r = 0.25\, \mathrm{km}$ and $0.5\, \mathrm{km}$  models, while the optical depth is almost
 smooth in the $\Delta r =1.0 \, \mathrm{km}$ model. 
If the spatial scale of self-gravity wakes is larger than the star
 projected radius, the fluctuation due to self-gravity wakes may be
 detected by occultation observations. 
 However, the observations show the smooth distribution.
Thus, the spatial scale of self-gravity wakes would be smaller than
 $1\,\mathrm{km}$, which leads to 
\begin{equation}
r_\mathrm{p} \lesssim
  13.7 \left(\frac{\rho_\mathrm{p}/\rho_\mathrm{C}}{0.5}\right)^{-1}
  \left(\frac{\tau}{0.38}\right)^{-1} \, \mathrm{m},
\label{eq:rp}
\end{equation}
 which is consistent with the estimate by \cite{Pan2016}.
In future observations, if we detect spatial variation of the
 optical depth, this will confirm the existence of self-gravity wakes and give a lower boundary for the particle size.

 If we consider the ring mass estimated by \cite{Pan2016}, we can give the lower boundary of the particle size. The surface density is estimated as $\mbox{few}\times 100\,\mathrm{g}\,\mathrm{cm}^{-2}$  from the apse alignment argument. 
From the ring formation criterion $\rho_\mathrm{p}/\rho_\mathrm{C}<0.52$, we give the following constraint
\begin{equation}
  r_\mathrm{p} \gtrsim 3.8 \left(\frac{\Sigma}{100\,\mathrm{g}\,\mathrm{cm}^{-2}} \right) \mathrm{m}.
\end{equation}

\subsection{Ring Lifetime}

The self-gravity wakes cause efficient radial diffusion in the inner
 ring. 
The effective viscosity of self-gravity wakes was obtained by $N$-body
 simulations \citep{Daisaka2001, Tanaka2003}, which is given as 
 $\nu_\mathrm{sgw} \simeq 26 \tilde r_\mathrm{H}^5 {G^2 \Sigma^2}/{\Omega^3}$.
Using this, we obtain the diffusion time of the inner ring as
\begin{equation}
t_\mathrm{inner} \equiv \frac{W^2}{\nu} \simeq
113 \left(\frac{r_\mathrm{p}}{1\,\mathrm{m}}\right)^{-2}
 \left(\frac{\rho_\mathrm{p}}{0.5\,\mathrm{g}\,\mathrm{cm}^{-3}}\right)^{-11/3}
 \mathrm{years},
\label{eq:tinner}
\end{equation}
 where we assume $W = 6.7\,\mathrm{km}$, $\tau = 0.38$, and
 $a =390.6 \, \mathrm{km}$.
This diffusion time is much shorter than in previous studies 
 \citep{Braga-Ribas2014, Pan2016} where the self-gravity wakes are not
 taken into account.

For the outer ring we estimate the diffusion time using the collisional viscosity,
 $\nu_\mathrm{col} \simeq r_\mathrm{p}^2 \Omega \tau$, which is
\begin{equation} 
t_\mathrm{outer} \simeq
 7.1 \times 10^4 \left(\frac{r_\mathrm{p}}{1\,\mathrm{m}} \right)^{-2}
 \mathrm{years},
\end{equation}
 where we assume $W = 3.5\,\mathrm{km}$, $\tau = 0.06$, and
 $a = 404.8 \, \mathrm{km}$.
This is roughly consistent with the results of previous studies
 \citep{Braga-Ribas2014, Pan2016}.

Equation (\ref{eq:tinner}) shows that the lifetime of the inner ring is
 much shorter than the typical dynamical lifetime of Centaurs,
 $\sim 10^{6}$ years, if the particle size is on the order of 1 m.
 If the rings were formed at the same time when Chariklo was scattered into the Centaur region, this significantly short ring lifetime would be inconsistent.
One possible solution to this contradiction is that Chariklo's rings are very young.
Another solution is that the ring consists of small particles, such as particles of less than $1\,\mathrm{cm}$.
It is also possible that the narrow ring is due to shepherding satellites,
 which will be discussed below \citep{Braga-Ribas2014}.

\section{Summary and Discussion \label{sec:summary}}

We performed global $N$-body simulations of Chariklo's rings and
 investigated their structure. 
We found that in order for Chariklo to host rings instead of particle aggregates its density should be larger than that of particles.
Under this condition, the self-gravity wakes inevitably develop in the
 inner ring independently of the particle size, while their spatial scale
 depends on the particle size. 
For m-sized ring particles, the timescale of ring viscous spreading due to
 the self-gravity wakes is on the order of $100\,\mathrm{years}$, which
 is much shorter than that estimated in previous studies \citep{Braga-Ribas2014, Pan2016}.

Our simulations predict that Chariklo is denser than the ring particles,
 and the ring particles are also less dense than ice, similar
to Saturn's rings.
In Saturn's rings, from comparisons of observations and $N$-body simulations, a particle density less than that of ice has been proposed \citep{Salo2004, Michikoshi2015}.
The higher density of Chariklo than that of the ring particles
 suggests that Chariklo may have a dense core.
Previous studies have hypothesized that the ring material originates from
 the stripped icy mantle of a differentiated body \citep{Hyodo2016}, which is consistent with our findings. 

The timescale of viscous ring spreading (Equation (\ref{eq:tinner}))
suggests three possibilities: a very young ring ($\sim 1\mbox{--}100\,\mathrm{years}$), 
smaller ring particles, or existence of shepherding satellites. If the ring is young, because a close
encounter with a giant planet is rare \citep{Araujo2016}, the ring
formation by the tidal interaction with giant planets may be
difficult. Then the other mechanisms, such as out gassing, would be 
preferable \citep{Pan2016}. If the ring particles are smaller than
$\sim 1$ cm, the inner ring can last longer than $10^6$
years. However, this small particle size is inconsistent with the
estimate from the apse alignment of the ring \citep{Pan2016}.
For the m-sized particles the existence of shepherding satellites is required to counteract the viscous ring spreading.
The minimum mass of the shepherding satellite depends on the
 distance from the inner ring edge $d$ and the particle size and density \citep{Goldreich1982}, 
\begin{equation}
M_\mathrm{s} \simeq 4.1 \left(\frac{\nu d^3}{\Omega a^5} \right)^{1/2} M_\mathrm{C}
= 1.3\times 10^{17} \left(\frac{d}{100\, \mathrm{km}} \right)^{3/2}
 \left(\frac{\rho_\mathrm{p}}{0.5\, \mathrm{g}\,\mathrm{cm}^{-3}} \right)^{11/6} 
  \left(\frac{r_\mathrm{p}}{1\, \mathrm{m}} \right) \mathrm{g},
\end{equation}
where we assumed the viscosity due to the self-gravity wakes, which is $\nu \propto \Sigma^2$.
The size of a satellite with mass $10^{17} \mathrm{g} $ is on the
 order of kilometers. 
Furthermore, the shepherding satellite hypothesis may be preferable because it could also explain the ring eccentricity.
 
The particle size is a key parameter for determining the dynamical
 property of Chariklo's rings.
However, the particle size has not yet been constrained observationally.
In future, if the self-gravity wakes are detected by higher
 resolution occultation observations, they will be able to provide a lower limit for the
 particle size.  

Our simulations suggest that the formation of self-gravity wakes is
a general process in dense narrow rings. For example, the Huygens
ringlet in the Cassini division of Saturn's ring and the $\epsilon$
ring of Uranus have a sufficiently high optical depth to form
self-gravity wakes. However, their spatial scale is far below the
observational limit today.

In this study, we assumed the circular ring, though the inner ring may have a finite eccentricity.
The effect of the eccentricity on the self-gravity wake dynamics has not been examined.
In the subsequent work, we plan to investigate its effect with more realistic ring models considering, for example, the size distribution of particles, the restitution coefficient that depends on the collisional velocity \citep{Bridges1984}, and shepherding satellites.

Numerical computations were carried out on ATERUI (Cray XC30) at the Center for Computational Astrophysics, National Astronomical Observatory of Japan.

\end{document}